# Atomically Resolved Acoustic Dynamics Coupled with Magnetic Order in a van der Waals Antiferromagnet

*Faran Zhou,\* Kyle Hwangbo, Sung Soo Ha, Xiao-Wei Zhang, Sae Hwan Chun, Jaeku Park, Intae Eom, Qianni Jiang, Zekai Yang, Marc Zajac, Sungwon Kim, Sungwook Choi, Zhaodong Chu, Kyoung Hun Oh, Yifan Su, Alfred Zong, Elton J. G. Santos, Ting Cao, Jiun-Haw Chu, Stephan O. Hruszkewycz, Nuh Gedik, Di Xiao, Hyunjung Kim, Xiaodong Xu, and Haidan Wen\**

F. Zhou, M. Zajac, Z. Chu, S. O. Hruszkewycz, H. Wen
Materials Science Division, Argonne National Laboratory, Lemont, Illinois 60439, USA
E-mail: nkzhoufaran@gmail.com; wen@anl.gov

K. Hwangbo, Q. Jiang, J.-H. Chu, D. Xiao, X. Xu
Department of Physics, University of Washington, Seattle, Washington 98195, USA

S. Ha, S. Kim, S. Choi, H. Kim
Center for Ultrafast Phase Transformation, Department of Physics, Sogang University, Seoul 04107, Korea

X.-W. Zhang, T. Cao, D. Xiao, X. Xu
Department of Materials Science and Engineering, University of Washington, Seattle, Washington 98195, USA

S. Chun, J. Park, I. Eom
Pohang Accelerator Laboratory, POSTECH, Pohang, Gyeongbuk 37673, Korea

Z. Yang, E. J. G. Santos
Institute for Condensed Matter and Complex Systems, School of Physics and Astronomy, The University of Edinburgh, Edinburgh EH9 3FD, United Kingdom

E. J. G. Santos
Donostia International Physics Center, Donostia-San Sebastian, 20018, Spain

K. Oh, Y. Su, N. Gedik
Department of Physics, Cambridge, Massachusetts Institute of Technology, Massachusetts 02139, USA

A. Zong
Departments of Physics and Applied Physics, Stanford University, Palo Alto, California 94305, USA

## Keywords

Magnetoelastic coupling, van der Waals magnets, coherent acoustic phonons, ultrafast x-ray diffraction

## Abstract

Magnetoelastic coupling in van der Waals (vdW) magnetic materials enables a unique interplay between the spin and lattice degrees of freedom. Characterizing the elastic responses with atomic and femtosecond resolution across the magnetic transition is essential for guiding the design of magnetically tunable actuators and strain-mediated spintronic devices. Here, ultrafast x-ray diffraction employed at a free-electron laser reveals that the atomic displacements, wave vectors, and dispersion relations of acoustic phonon modes in a vdW antiferromagnet $FePS_3$ are coupled with the magnetic order, by tracking both in-plane and out-of-plane Bragg peaks upon optical excitation across the Néel temperature ($T_N$). One transverse mode shows that a quasi-out-of-plane atomic displacement undergoes a significant directional change across $T_N$. Its quasi-in-plane wave vector is derived by the comparison between the measured sound velocity and the first-principles calculations. The other transverse mode is an interlayer shear acoustic mode whose amplitude is strongly enhanced in the antiferromagnetic phase, exhibiting eight times stronger amplitude than the longitudinal acoustic mode below $T_N$. The atomically resolved characterization of acoustic phonon dynamics that couple with magnetic ordering opens opportunities for harnessing unique magnetoelastic coupling in vdW magnets on ultrafast timescales.

# 1. Introduction

Magnetoelastic coupling in magnetic materials holds great promise for sensing, actuation, and energy harvesting applications.[1–4] Van der Waals (vdW) magnets are particularly compelling materials for these applications due to their diverse magnetic states, remarkable mechanical flexibility, and strong magnetoelastic coupling.[5–9] Coherent acoustic phonons play critical roles in governing nanoscale energy and momentum transport. However, the characteristics of acoustic phonons, such as their atomic displacements, wave vectors, and dispersion, are challenging to study using conventional optical techniques[10] that have negligible momentum transfer. By contrast, ultrafast x-ray diffraction enables momentum-resolved tracking of acoustic wave propagation with atomic sensitivity.[11] Quantitative and direct measurements of coherent acoustic phonons and their relations with magnetic order offer insights into the unique magnetoelastic coupling in vdW magnets.

FePS$_3$ is featured in this study as an exemplary vdW magnet because it exhibits strong spin-lattice coupling. In this material, both the magnetic and structural phase transitions occur at the Néel temperature ($T_N$ = 117 K).[12,13] Below $T_N$, FePS$_3$ hosts a zigzag in-plane antiferromagnetic order within each honeycomb layer, as well as out-of-plane antiferromagnetic order between layers that are stacked in a monoclinic crystal structure.[14] The strong spin-lattice coupling is evidenced in several observations, including the mechanical resonance sensitive to spin orders,[4] demagnetization-coupled optical phonons,[15] magnon-phonon hybridization,[16] coupled critical slowing down,[5] spin-mediated shear oscillations,[17] and exotic THz field-induced metastable ferromagnetic states.[18] Previous ultrafast electron scattering and imaging studies[17,19] have focused on in-plane Bragg peaks, thus mainly sensitive to in-plane atomic motions, leaving the coherent acoustic phonons with out-of-plane atomic displacements largely unexplored.

Here, we report atomically resolved characterization of three coherent acoustic modes in the vdW antiferromagnet FePS$_3$ using ultrafast x-ray diffraction. By tracking the dynamics of both in-plane and out-of-plane Bragg peaks following photoexcitation above the band gap, we directly measure the atomic displacements and momentum-energy dispersions of these three modes. The first mode with the lowest frequency exhibits atomic displacements along the out-of-plane direction while its wave vector is aligned with a quasi-in-plane direction. This mode is a quasi-transverse mode that propagates nearly perpendicular to the transient temperature gradient. Notably, its atomic displacement vector rotates by ~20° and exhibits a ~5% change in sound velocity across $T_N$, indicating strong coupling to the magnetic order. The second mode with an intermediate frequency is identified as an interlayer shear acoustic mode with in-plane atomic displacements and an out-of-plane propagation direction. Its amplitude is significantly suppressed above $T_N$, which is consistent with earlier reports.[17,19] The third mode with the highest frequency is a longitudinal interlayer breathing mode,[20] characterized by both the atomic displacement and wave vector oriented along the out-of-plane direction. Below $T_N$, the oscillation amplitude of the interlayer shear mode is eight times larger than that of the longitudinal interlayer breathing mode, a ratio that exceeds the largest previously reported value of six in the multiferroic perovskite system BiFeO$_3$.[21]

## 2. Results and Discussions

We performed 400-nm-optical-pump, hard-x-ray-diffraction-probe measurements of nanometer-thick FePS$_3$ flakes at the Femtosecond X-ray Scattering (FXS) endstation of the Pohang Accelerator Laboratory X-ray Free-Electron Laser (PAL-XFEL), as illustrated in Figure 1a. Multiple flakes with uniform thickness ranging from 40 nm to 200 nm were studied. The 100-fs optical pump pulses were focused to a spot size with a spatial extent of ~300 μm × 1500 μm in full width at half maximum (FWHM) on the sample, which was larger than the probed region to ensure homogeneous excitation. The pump photon energy of 3.1 eV is above the band gap (1.7 eV) of FePS$_3$.[22] The 12.0-keV x-ray pulses, operating at a repetition rate of 60 Hz, were focused to a ~15 μm × 75 μm (FWHM) spot on the sample surface. At an incident optical fluence of 2.6 mJ cm$^{-2}$, the top surface of the sample absorbed 0.16 photons per unit cell, leading to a transient lattice temperature rise of ~120 K (see Section 1 in Supporting Information) after the electron-phonon coupling. The transiently established temperature gradient along the sample depth, together with the released structural instabilities associated with ultrafast demagnetization,[17] leads to the coherent generation of acoustic phonons that are captured in real time using ultrafast x-ray diffraction.

We focused our measurements on the shifts of 002 and $\bar{2}$02 Bragg peaks, which are sensitive to the atomic displacements in the *ac*-plane (Figure 1b), as the laser-induced structural changes in FePS$_3$ primarily occur within the *ac*-plane.[5] Upon optical excitation, the 002 Bragg peak was observed to shift toward the -**c*** direction, as captured by a two-dimensional x-ray area detector, indicating a laser-induced interlayer expansion (Figure 1c). This observation, which was quantified by analyzing the peak shift in the two-dimensional reciprocal space cut accessible to the detector, is consistent with the three-dimensional reciprocal space mapping of the peak shift extracted from the time-resolved rocking curve measurements (Figure S1). In addition to the step-like peak shift, oscillatory modulations were superimposed on the 002 peak dynamics. A fast Fourier transform (FFT) of the time-dependent 002 peak shift revealed a strong frequency component at 3.5 GHz and a weak one at 11.1 GHz (Figure 1d), indicating the excitation of at least two coherent phonon modes. Temperature-dependent FFT amplitudes showed that both modes persist across $T_N$. The mode at 3.5 GHz dominates the oscillation, with an FFT amplitude more than four times larger than that of the 11.1-GHz mode. On the other hand, for the $\bar{2}$02 Bragg peak dynamics (Figure 1e), the most prominent temperature-dependent feature was a reversal in the sign of the peak shift ($\Delta q_z$): negative above $T_N$ and positive below $T_N$. This behavior is consistent with an interlayer shear motion, attributed to a sudden decrease in the monoclinic angle $\beta$ from its equilibrium value of 107.2°, triggered by the melting of the antiferromagnetic order as reported previously.[5,17] In addition, the oscillatory signal of $\bar{2}$02 peak exhibited different periods and the oscillation phase underwent a $\pi$ shift across $T_N$. The oscillation amplitude was shown by the FFT spectra (Figure 1f), which revealed the emergence of a new mode with a large FFT amplitude at 4.6 GHz below $T_N$. These coherent acoustic modes were consistently observed across multiple samples, despite the frequency variations due to differences in sample thickness. The linear dependence of the oscillation periods on the sample thickness (Figure S2) confirmed that all three modes we observed were neither

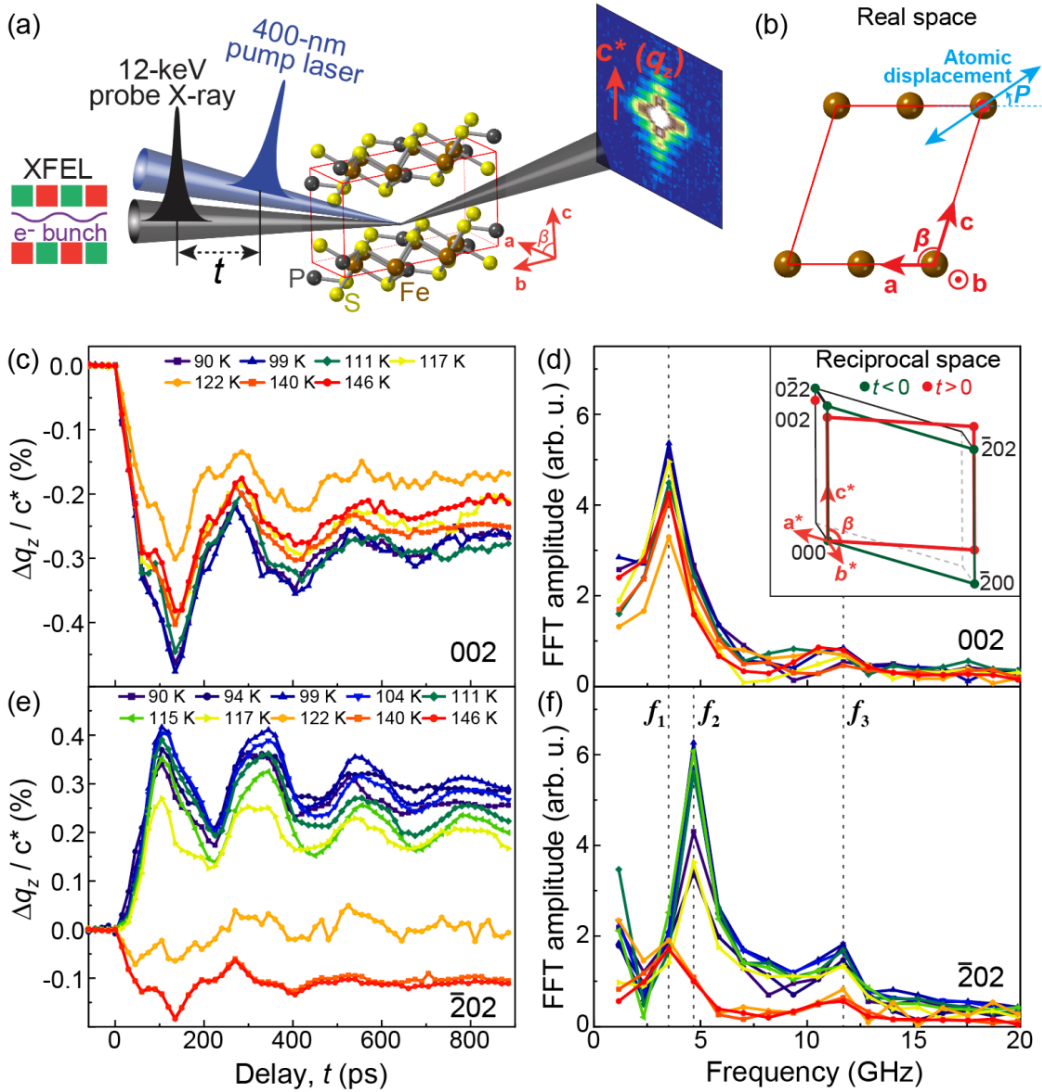

**Figure 1.** Schematic experimental setup and time-dependent Bragg peak shifts. (a) Schematic of the ultrafast x-ray diffraction experimental setup using XFEL (x-ray free-electron laser). (b) Real-space structure viewed from the *b*-axis. Only Fe atoms are shown for clarity. The double blue arrows indicate the atomic vibration and polarization angle *P* relative to the *a*-axis. (c) Time-dependent center-of-mass shift ($\Delta q_z$) of the 002 Bragg peak along the $\mathbf{c^*}$ direction at various temperatures. (d) Fast Fourier transform (FFT) of the time-resolved data in (c). Inset, schematic of the reciprocal lattice of FePS$_3$. Green and red symbols represent the peak positions before and after laser excitation, respectively. (e) Time-dependent center-of-mass shift ($\Delta q_z$) of the $\bar{2}02$ Bragg peak along the $\mathbf{c^*}$

direction at various temperatures. (f) FFT of the time-resolved data in (e). The vertical dashed lines indicate the frequencies of the three phonon modes, labeled as $f_1$, $f_2$, and $f_3$.

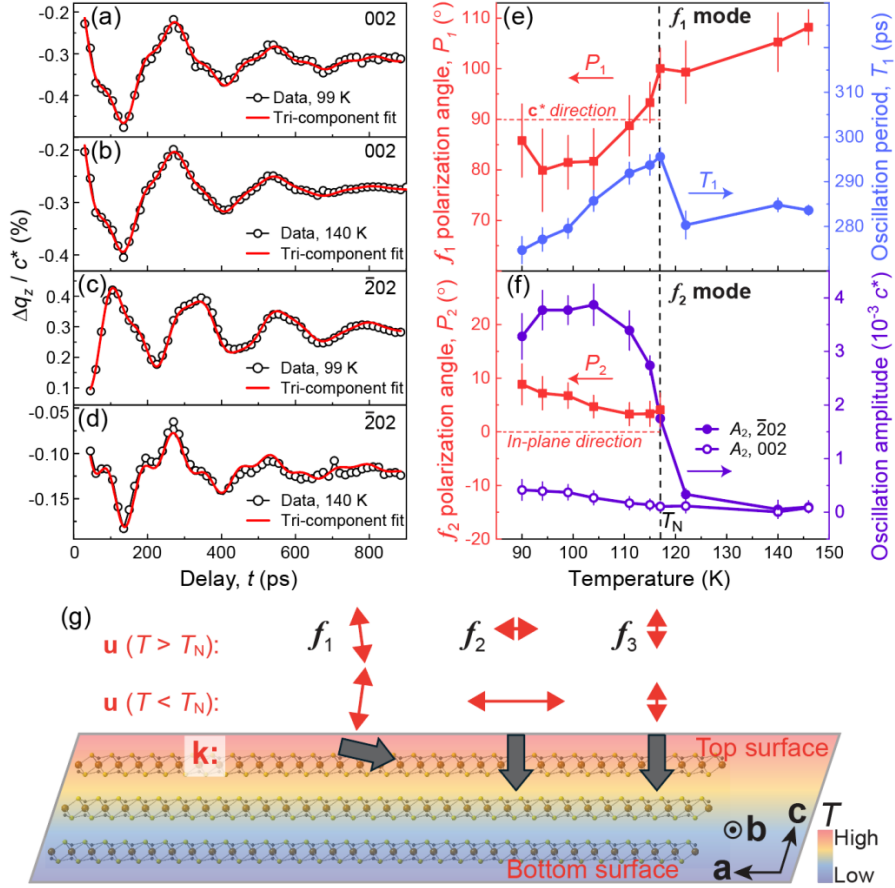

**Figure 2.** Time-domain fitting of the oscillatory dynamics. (a)-(b) Oscillatory dynamics of the 002 Bragg peak measured at 99 K (a) and 140 K (b). To emphasize the oscillatory behavior, the data points at negative time delays are omitted. Red solid lines are fits using a three-component damped sinusoidal model. (c)-(d) Same as (a)-(b), but for the $\bar{2}02$ Bragg peak. (e) Temperature-dependent atomic polarization angle $P_1$ relative to the $a$-axis direction and oscillation period $T_1$ for the $f_1$ mode. The definition of polarization angle is illustrated in Figure 1b. (f) Temperature-dependent atomic polarization angle $P_2$ relative to the $a$-axis direction and oscillation amplitudes $A_2$ of two peaks for the $f_2$ mode. The polarization angle above $T_N$ is omitted due to the disappearance of the $f_2$ mode. The error bars in (e) and (f) are errors propagated from the standard deviation of the time-domain fitting. (g) Schematic drawing of the atomic displacement vector **u** and the wave vector **k** direction for the three modes. The displacement vector **u** is based on the time-domain fittings, while the wave vector **k** will be discussed in later sections. The parallogram represents the sample flake that establishes a temperature gradient after optical excitation.

optical phonons nor magnons, but acoustic phonons whose wave vectors **k** have an out-of-plane component. For clarity, we labeled the three acoustic phonon modes as $f_1$, $f_2$, and $f_3$, corresponding to the 3.5 GHz, 4.6 GHz, and 11.1 GHz modes, respectively.

When multiple oscillatory signals have similar frequencies and finite lifetimes, as in the case of the $f_1$ and $f_2$ modes, it is difficult to determine their amplitudes and frequencies precisely based on the FFT analysis alone. To complement the FFT analysis, we fitted the data by a function that is composed of three sinusoidal functions with exponentially decaying amplitudes: $F(t) = \sum_{i=1,2,3}(A_i e^{-\frac{t}{\tau_i}} \cos(2\pi f_i t + \varphi_i)) + cc$, where $A_i$, $\tau_i$, $f_i$, and $\varphi_i$ are the amplitude, decay constant, frequency, and phase of the $i$-th ($i$ = 1, 2, 3) mode, respectively, and $cc$ is a constant offset (See Section 2 in Supporting Information for details). Figures 2a-2d show the fittings for the 002 and $\bar{2}02$ peak, respectively, with data from below $T_N$ in the top panels and above $T_N$ in the bottom panels. The temperature-dependent fitting results were summarized in Figure S3, S4. For the 002 peak at all temperatures, the average amplitude of the $f_1$ mode was four times larger than the $f_3$ mode and ten times larger than that of $f_2$ mode (Figure S3a), indicating a dominant $f_1$ component and a negligible $f_2$ component measured in the 002 peak. For the $\bar{2}02$ peak above $T_N$, the fitting results were similar to those of the 002 peak. By contrast, for the $\bar{2}02$ peak below $T_N$, the $f_2$ mode dominates the oscillation with the largest amplitude (Figure S3a) and exhibits $\pi$ out of phase (Figure S4a) compared to the other two modes.

Based on the reciprocal space geometry (inset of Figure 1d) and the oscillation amplitudes, we were able to extract the polarization angles (See Section 3 in Supporting Information), i.e., the direction of atomic displacement vectors **u**, for all three modes. The shift of the 002 peak along **c*** ($\Delta q_z$) is a result of the out-of-plane atomic displacements, while the shift of $\bar{2}02$ peak is determined by both the in-plane (along the $a$-axis) and the out-of-plane atomic displacements. Since both peaks share the same $L$-index, any difference in their $\Delta q_z$ shift (Figure S3b) reflects the changes in the monoclinic tilt angle $\beta$ and in-plane atomic displacements. As shown in Figure 2e, the polarization of the $f_1$ mode was primarily along the **c*** direction below $T_N$ and rotated by about 20° above $T_N$, which likely originates from the strong magnetoelastic coupling. Crossing $T_N$, the change in spin exchange interactions leads to a renormalization of the inter-atomic force constants and elastic tensor,[4,23] which subsequently leads to a rotation of the displacement polarization. In addition, the oscillation period (Figure 2e) of the $f_1$ mode changed by 5% across $T_N$, further showing the effect of magnetic ordering on the elastic properties associated with the $f_1$ mode. For the $f_2$ mode (Figure 2f), its amplitude is the largest among all three modes below $T_N$, but is suppressed to nearly zero above $T_N$. In contrast to the out-of-plane atomic displacement for the $f_1$ and $f_3$ modes (Figure S3c), the polarization of the $f_2$ mode is mostly in the in-plane direction. We estimate the systematic error is 5° for the absolution value of the measured polarization angles. The atomic displacements and wave vectors of all three modes are summarized in Figure 2g, which is further supported by the dispersion analysis and first-principles calculations detailed below.

Besides the atomic displacements, ultrafast x-ray diffraction offers momentum-resolved dynamics that enable the direct characterization of the acoustic phonon dispersions. We analyzed the intensity oscillations of the Kiessig fringes on the two-dimensional x-ray area detector (Figure 3a) that emanate from the Bragg peak due to the finite-thickness effects of the crystal and cover a span of momentum. To extend the accessible momentum range afforded by these fringes, we decreased the sample $\theta$ angle by 0.1° so that the fringes were stronger on the lower-$q_z$ side. We note that the frequency we measured here in the dispersion relation is not the second harmonic of the acoustic waves as measured by the time-domain inelastic x-ray scattering technique, because the dominant dynamical signal in our measurement is still near the Bragg peak rather than the diffuse scattering.[24] After optical excitation, different Kiessig fringes exhibited distinct oscillation periods, as highlighted by the red dashed curves in Figure 3b. With the correction that removed the contribution from the strain pulse propagation (Figure S5), we performed FFT on the time evolution of each momentum point on the shift-corrected image to obtain the dispersion map shown in Figure 3c. A prominent linear dispersion in Figure 3c confirms the acoustic nature of this mode. By fitting the slope of the dispersion (Figure S6a), we extract a sound velocity of $3561 \pm 23$ m s$^{-1}$. This value is lower than the calculated value of $4195 \pm 19$ m s$^{-1}$ by $v = \frac{2d}{T}$, where $d = 195.8$ nm is the flake thickness as measured by the Kiessig fringe and $T = 93.3$ ps is the fitted oscillation period. This discrepancy might be due to the difference in the effective propagation distance of the acoustic wave in the two measurements, as detailed in Section 4 of the Supporting Information. Nevertheless, the measured sound velocity of the $f_3$ mode is highest among all the measured modes and agrees with the calculated longitudinal acoustic velocity (the value of the purple curve at 90° in Figure 4a). Therefore, the mode $f_3$ is ascribed as a quasi-longitudinal acoustic mode, i.e., the interlayer breathing mode.

The dispersion of the $\bar{2}02$ peak below $T_N$ is shown in Figure 3d-3f. The sound velocity from linear dispersion fitting is $1668 \pm 18$ m s$^{-1}$ (Figure S6b), in agreement with the sound velocity of the $f_2$ mode ($1709 \pm 15$ m s$^{-1}$) estimated by the oscillation of the $\bar{2}02$ peak shift. Furthermore, the dispersion of this mode was prominent below $T_N$ but strongly suppressed above $T_N$ (Figure 3g-i), in agreement with the temperature dependence of the mode amplitude measured by the dynamics of the peak shift (Figures 2c-2d) as well as by the previous ultrafast electron diffraction and microscopy measurements.[17,19] Therefore, the $f_2$ mode is an interlayer shear acoustic mode that propagates along the out-of-plane direction with an in-plane atomic displacement. We note that the sound velocity measured by this work is higher than the value measured by ultrafast electron diffraction, [17] which may be related to the different sample environment as detailed in Section 4 of the Supporting Information.

For the lowest-frequency $f_1$ mode, its $\Delta q_z$ oscillation (Figure 1c and 2a) dominates the 002 peak dynamics, with an amplitude four times larger than the interlayer breathing $f_3$ mode. However, in the dispersion map shown in Figure 3c, only the $f_3$ mode is prominent, while the $f_1$ mode is not resolvable along the expected dispersion trajectory indicated by the red rectangle. This absence is consistent with the interpretation that the wave vector of the $f_1$ mode is mainly in-plane with little projection along the **c*** direction, so that its dispersion is not discernible along the **c*** direction. The wave vector of the $f_1$ mode cannot be purely in-plane either, because the oscillation period of the $f_1$ mode scales linearly with the sample thickness (Figure S2), implying the presence of an out-of-plane component in its wave vector. Therefore, the wave vector of the $f_1$ mode is tilted significantly away from the sample surface normal, with a dominant in-plane and a small out-of-plane component, as schematically shown in Figure 2g.

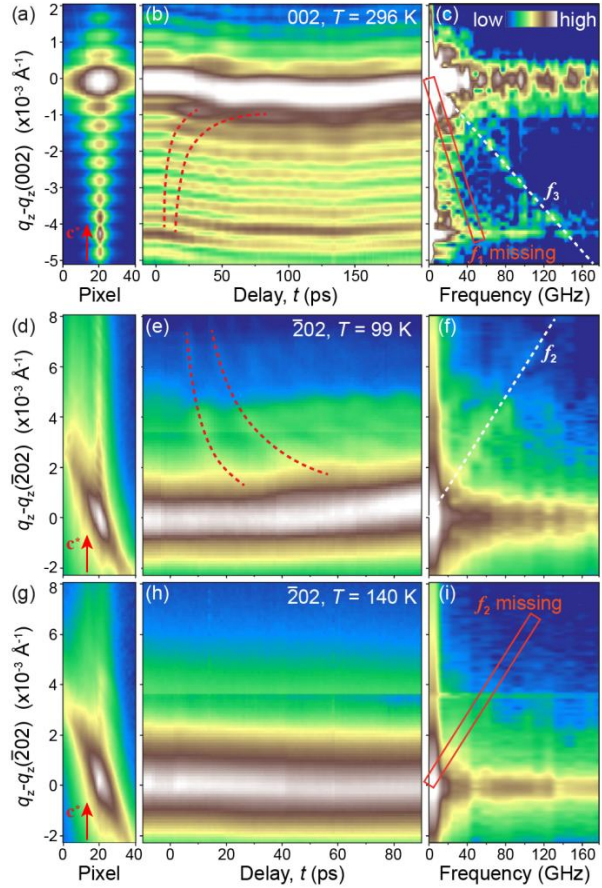

**Figure 3.** Momentum-resolved dynamics and phonon dispersions. (a) Diffraction image of the 002 Bragg peak displayed in a logarithmic scale. The central peak at $q_z - q_z(002) = 0$ is the 002 peak, and the side peaks along $q_z$ are the Kiessig fringes. (b) Momentum-resolved dynamics of the 002 peak at room temperature. Red dashed lines are visual guides indicating the $q_z$-dependent oscillation, the same as in (e). (c) FFT spectra of the data in (b). The white dashed line highlights the dispersion of the longitudinal acoustic mode ($f_3$), and the red rectangle indicates the expected dispersion for the $f_1$ mode. (d) Diffraction image of the $\bar{2}02$ Bragg peak displayed on a logarithmic scale. (e) Momentum-resolved dynamics of the $\bar{2}02$ peak at 99 K. (f) FFT spectra of the data in (e). The white dashed line traces the dispersions of the $f_2$ branch. (g)-(i) Same as (d)-(f), but measured at 140 K. The red rectangle indicates suppression of the $f_2$ mode above $T_N$.

To further assess the wave vector of the $f_1$ mode, we compare the measured sound velocity with the theoretical values. We performed first-principles calculations of FePS$_3$ and extracted sound velocities in all three dimensions. The sound velocities of three acoustic branches were plotted in Figure 4a along each wave vector direction in the $ac$-plane. The theory calculations were first verified by comparing with the experimental sound velocities of the $f_2$ and $f_3$ modes. The theory values for the two modes (marked in the vertical axis in Figure 4a) were 1715 and 3332 m s$^{-1}$, respectively, in agreement with the measured sound velocities. For the $f_1$ mode, since its wave vector is neither purely in-plane nor out of plane as discussed earlier, we need to search for a point on the TA1 curve between 0 and 90 degrees that matches the measured sound velocity. TA1 branch is chosen instead of TA2 because the phonon polarization of TA1 branch matches that of the $f_1$ mode. A horizontal dashed line with a height of the measured velocity $V_{1p}$ along the c* direction is drawn to find its crossing point with the TA1 branch. The origin to this crossing point defines the wave vector of the $f_1$ mode, which is oriented at 30º with respect to the in-plane direction. The length of the vector represents the velocity along this direction, whose value can be calculated as $V_{1p} / \sin 30° = 2591 \pm 118$ m s$^{-1}$.

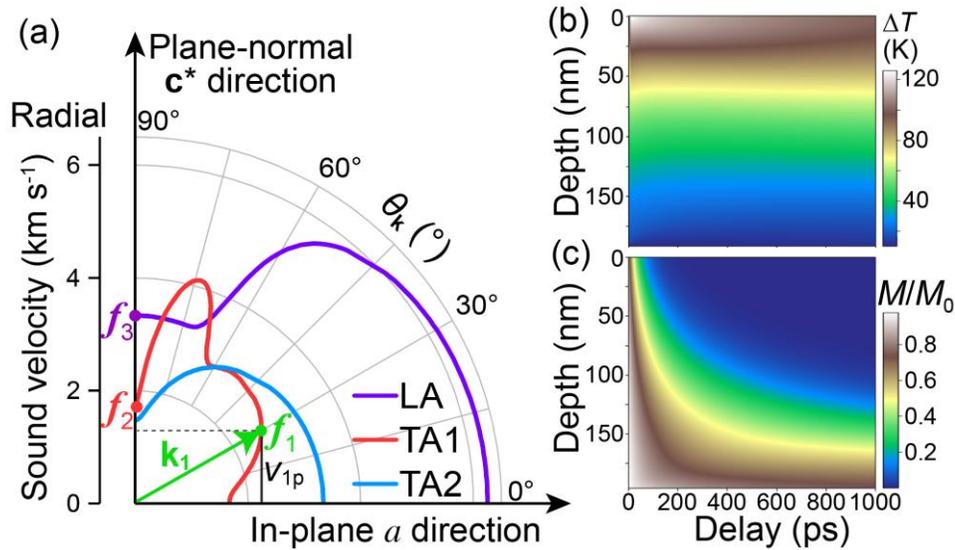

**Figure 4.** Sound velocity and thermal transport calculations. (a) Polar plot of the sound velocities for three acoustic branches as a function of the wave vector directions ($\theta_k$) in the $ac$-plane. $\theta_k = 0°$ in the graph represents the $a$-axis, while $\theta_k = 90°$ corresponds to the plane-normal $c*$ direction. Calculation details are in the Experimental Section. LA: longitudinal acoustic. TA: transverse acoustic. The horizontal dashed line represents the experimentally measured $f_1$ sound velocity, which is a projected velocity along $c*$ labelled $v_{1p}$. The crossing point between the dashed line and the red line TA1 branch represents the $f_1$ mode with wave vector labelled $\mathbf{k}_1$. (b) Depth-dependent sample temperature rise ($\Delta T$) calculated by a microscopic three-temperature model, detailed in Section 5 of the Supporting Information. (c) Depth-dependent demagnetization dynamics normalized to the initial magnetization $M_0$.

## 3. Discussion

It is unusual that the wave vector of the $f_1$ mode is mostly in-plane. In previous studies of photoexcited coherent acoustic phonons, the phonon propagation direction, regardless of a longitudinal or transverse mode, has always been expected to be along the surface normal direction.[11,25–27] This is because photoexcitation of the sample's top surface creates a transient temperature gradient and thus thermal stress along the sample depth direction. While it is common to generate in-plane or surface acoustic waves[27] from the laser-pumped region propagating towards unpumped areas due to the in-plane thermal stress, our measurements focused on the homogeneously excited region within the pump laser spot. In this case, the pump laser spot size is much larger than the probe spot size, ensuring homogeneous excitation on the sample's top surface of the probed region. In the previous reports of in-plane travelling acoustic phonons, analogous to the $f_1$ mode, the coherent phonons have always been launched from the sample edges or defects.[19,29–31] In our study, the $f_1$ mode cannot originate from edges because the probed region is at least 10 μm away from the sample edges. The acoustic wave launched at the edges would result in a delayed onset time of nanosecond timescale in the dynamics, which we did not observe. On the other hand, it is possible that the $f_1$ mode is launched from nanoscale structural defects such as step edges, wrinkles, or surface roughness,[32] which are commonly present in vdW materials. These structural defects in FePS$_3$ were confirmed by the previous ultrafast electron microscopy investigations.[17,19]

Our results show that the wave vector of the $f_1$ mode is directed in the quasi-in-plane direction, indicating a transverse thermal flow perpendicular to the temperature gradient. Strong magnetoelastic coupling in FePS$_3$ has been proposed to induce hybridization between magnon and phonon bands, which, combined with time-reversal symmetry breaking, gives rise to a nonzero Berry curvature and the thermal Hall effect.[33] The predicted band hybridization between magnons and optical phonons occurs near 14 meV (~3.4 THz). At the energy range of acoustic phonons observed in this study, the acoustic phonon and magnon hybridization are not expected in antiferromagnetic materials due to their energy and momentum mismatch. Nevertheless, our temperature-dependent characterization of acoustic phonons clearly indicates that the magnetic order plays an important role in governing the dynamics of $f_1$ mode. The efficient transverse thermal flow carried by $f_1$ mode presents a promising avenue for future phononic and elastic actuation applications. Conversely, the strong magnetoelastic coupling demonstrated here could be potentially harnessed for coherent control of magnetism via acoustic phonon excitation.

The interlayer shear acoustic mode $f_2$ has been reported in the previous electron scattering studies.[17,19] Here, with ultrafast x-ray diffraction measurements, we provided direct evidence for both the atomic displacements and phonon dispersions across $T_N$. Apart from the $f_1$ and $f_2$ modes, the interlayer breathing mode $f_3$ was reported and comprehensively characterized for the first time. These results allow for quantitative comparisons between these modes. Specifically, the ratio of oscillation amplitudes $A_2/A_3$ reveals that the interlayer shear mode exhibits an amplitude eight times larger than that of the interlayer breathing mode. This striking contrast underscores the presence of giant shear instabilities strongly coupled to the zigzag antiferromagnetic order, which has been demonstrated to be effective for ultrafast demagnetization-driven shear oscillations.[17,19] Below $T_N$, ultrafast demagnetization alters the spin exchange interactions, which subsequently modifies the inter-atomic potential and thus shifts the minimum-energy atomic positions. Above $T_N$, photo excitation does not change the spin exchange interactions, resulting in diminishing amplitude for the $f_2$ mode.

To corroborate the atomic dynamics measured by ultrafast x-ray diffraction, we performed thermal and spin transport simulations using a microscopic three-temperature model (M3TM, see Section 5 in Supporting Information for details). The simulation captures the evolution of the lattice and spin temperatures (Figures 4b and 4c) across the sample, in which thermal gradient persists beyond 1 ns. Since the sample is in a mixed state of antiferromagnet and paramagnet after excitation in thick samples, but mostly a paramagnetic state in thinner samples, the speed of sound for the interlayer shear mode is expected to be higher in the antiferromagnetic state. As the magnetization state of the sample still evolves in the thick sample during the measured time window (Figure 4c), the resulting acoustic transport is thus more complicated to model. This complication may partially explain why the sound velocity of the interlayer shear ($f_2$) mode we measured in this work is 33% larger than that measured in significantly thinner samples by ultrafast electron diffraction and microscopy.[17] Other factors may also contribute to this discrepancy (Section 4 in Supporting Information).

## 4. Conclusion

We have investigated the coherent acoustic phonon dynamics in the exemplary vdW antiferromagnet FePS$_3$ across its magnetic ordering temperature using ultrafast time-resolved x-ray diffraction. In addition to the previous electron scattering studies,[17,19] which primarily probed in-plane atomic displacements, our measurements revealed previously unobserved coherent phonon modes ($f_1$ and $f_3$) exhibiting out-of-plane atomic displacements and unconventional propagation characteristics. Notably, direct characterization of the interlayer shear mode ($f_2$) and the interlayer breathing mode ($f_3$) yields a giant shear-to-longitudinal oscillation amplitude ratio of up to eight, which highlights the shear degrees of freedom that couple to magnetic order. Further, the $f_1$ mode exhibits a transverse propagation direction perpendicular to the transient temperature gradient and a 20° rotation in its atomic polarization vector at $T_N$. This observation suggests strong magnetoelastic coupling and may have connections to the theoretically predicted nonzero Berry curvature-induced topological transverse transport phenomenon in FePS$_3$.[33] Together, the two transverse modes observed here thus provide a unique platform for exploring magnetoelastic coupling and engineering unconventional phonon transport in future phononic and spintronic devices.

## 5. Experimental Section

*Sample growth and sample preparation*: Single crystals of FePS$_3$ were synthesized by the chemical vapour transport method using iodine as the transport agent. Stoichiometric amounts of iron powder (99.998%), phosphorus powder (98.9%) and sulfur pieces (99.9995%) were mixed with iodine (1 mg cm$^{-3}$) and sealed in quartz tubes (10 cm in length) under high vacuum. The tubes were placed in a horizontal one-zone tube furnace with the charge near the center of the furnace. Sizeable crystals (10 × 10 × 0.5 mm$^3$) were obtained after gradually heating the precursor up to 750 °C, dwelling for a week and cooling down to room temperature. Ultrathin crystals were obtained by mechanical exfoliation, and thin flakes were transferred to sapphire substrates. The layer thickness was measured by atomic force microscopy as well as by measuring the Kiessig fringes in situ.

*Ultrafast X-ray diffraction experiment*: The ultrafast x-ray diffraction experiments were performed at the Femtosecond X-ray Scattering (FXS) endstation of the Pohang Accelerator Laboratory X-ray Free-Electron Laser (PAL-XFEL). The 400 nm pump laser pulse was derived by doubling the fundamental wavelength of an amplified Ti:Sapphire laser. The pump beam has a 10º-crossing angle relative to the x-ray probe beam. At the scattering geometry of the 002 Bragg peak, the pump laser footprint on the sample was 300 μm × 1500 μm (FWHM), with an incident fluence up to 4 mJ cm$^{-2}$ for the experiment. The repetition rates of the pump laser and the probe x-ray were 30 Hz and 60 Hz, respectively. The x-ray pulse with a duration of 40 fs was monochromatized to an energy of 12 keV and focused to a size of 15 μm × 75 μm (FWHM) on the sample by a pair of Kirkpatrick–Baez mirrors. The x-ray footprint is aligned at the center of the sample (~100 μm in lateral size, see Figure S1a), which makes the probe beam more than 10 μm from the sample edge as quoted in the Discussion section. The sample was mounted on a four-circle diffractometer (Huber GmbH.) and cooled down by a helium cryoblower (Oxford Instruments). A temperature sensor was mounted right underneath the sample for monitoring the sample temperature. An area detector (Jungfrau 0.5M) was used to record the shot-by-shot diffraction pattern. Regarding the absence of delayed onset dynamics in FePS$_3$, the time-zero was independently calibrated by a standard bithmuth sample with femtosecond precision.

*First-principles calculations of sound velocity*: First-principles density functional theory (DFT) calculations were carried out for FePS$_3$ in the zigzag antiferromagnetic configuration using the VASP package.[34] The local density approximation (LDA) functional was employed, with a Hubbard U correction applied to the Fe 3d orbitals. The correction was treated using the Dudarev approach with an effective U–J value of 3.5 eV.[35] A Γ-centered 6×3×3 k-point mesh was used for Brillouin zone sampling, and the plane-wave basis set was truncated at a kinetic energy cut-off of 500 eV. Structural relaxation was performed until both the lattice parameters and atomic positions were fully optimized. The elastic moduli tensor was then calculated using the finite difference method. Finally, sound velocities were derived from the computed elastic constants using the Christoffel package.[36]

*Microscopic three-temperature Model (M3TM):* Our theoretical approach to understanding ultrafast magnetization dynamics is based on the microscopic three-temperature model (M3TM).[37–41] In this framework, the ultrafast laser pulse $S(z,t)$ interacts with the electronic subsystem in the electric dipole approximation, where electrons near the Fermi energy get excited above the band gap. Upon fast thermalization of the electron system through Coulomb scattering, the laser pulse causes a substantial elevation in the electronic temperature $T_e$. Electron-phonon scattering processes allow for the exchange of energy and equilibration of $T_e$ and the phonon bath, characterized by its temperature $T_p$. Depth-dependent laser absorption leads to temperature gradients within the sample that cause diffusion processes in both baths. The energetic cost of demagnetization gets compensated by the electronic system. Additional details are in Section 5 of the Supporting Information.

## Acknowledgements

This work was primarily supported by the U.S. Department of Energy, Office of Science, Basic Energy Sciences, Materials Sciences and Engineering Division, under Award No.DE-SC0012509 (Experimental design, sample preparation, data collection and analysis, theory, and manuscript preparation by F.Z., K.H., Q.Z., C.W., L.S., A.Z., Y.S., N.G., X.X., D.X., and H.W.). Part of data collection by M. Z., S.H., and H.W. was supported by the U.S. Department of Energy, Office of Science, Basic Energy Sciences, Materials Sciences and Engineering Division. The time-resolved X-ray diffraction experiment was performed using the FXS instrument at PAL-XFEL (Proposal No. 2022-1st-XSS-028, 2022-2nd-XSS-027) funded by the Ministry of Science and ICT of Korea. Materials synthesis is supported by the University of Washington Molecular Engineering Materials Center, an NSF Materials Research Science and Engineering Center (Grant No. DMR-1719797 and DMR-2308979). S.S.H., S.K., S.C., and H.K. acknowledge the support by the National Research Foundation from the Ministry of Science and ICT of Korea (RS-2021-NR059920). E.J.G.S. acknowledges computational resources through CIRRUS Tier-2 HPC Service (ec131 Cirrus Project) at EPCC (http://www.cirrus.ac.uk) funded by the University of Edinburgh and EPSRC (EP/P020267/1); and ARCHER2 UK National Supercomputing Service via the UKCP consortium (Project e89) funded by EPSRC grant ref EP/X035891/1. E.J.G.S. acknowledges the EPSRC Open Fellowship (EP/T021578/1) and the Donostia International Physics Center for funding support. A.Z. acknowledges the support from the U.S. Department of Energy, Office of Basic Energy Sciences under award, No. DE-SC0026202.

## Conflict of Interest

The authors declare no conflict of interest.

## Author Contributions

F.Z., S.S.H., S.H.C., J.P., I.E., M.Z., S.K., S.C., Z.C., K.O., Y.S., A.Z., S.O.H., N.G., H.K., X.X., and H.W. performed ultrafast X-ray diffraction experiments. F. Z. and H.W. performed data analysis and interpreted the data. Q.J. grew the single crystals under the supervision of J.-H.C. K.H. prepared the samples supervised by X.X. X.Z. T.C, and D.X. did first-principles calculations. Z.Y. and E.J.G.S. simulated the time-dependent temperature profile. F.Z. and H.W. wrote the manuscript with input from all authors. H.W. conceived and supervised the project.

## Data Availability Statement

The data that support the findings of this study are available from the corresponding author upon reasonable request.

# Supporting Information

## Atomically Resolved Acoustic Dynamics Coupled with Magnetic order in a van der Waals antiferromagnet

*Faran Zhou,\* Kyle Hwangbo, Sung Soo Ha, Xiao-Wei Zhang, Sae Hwan Chun, Jaeku Park, Intae Eom, Qianni Jiang, Zekai Yang, Marc Zajac, Sungwon Kim, Sungwook Choi, Zhaodong Chu, Kyoung Hun Oh, Yifan Su, Alfred Zong, Elton J. G. Santos, Ting Cao, Jiun-Haw Chu, Stephan O. Hruszkewycz, Nuh Gedik, Di Xiao, Hyunjung Kim, Xiaodong Xu, and Haidan Wen\**

F. Zhou, M. Zajac, Z. Chu, S. O. Hruszkewycz, H. Wen
Materials Science Division, Argonne National Laboratory, Lemont, Illinois 60439, USA
E-mail: nkzhoufaran@gmail.com; wen@anl.gov

K. Hwangbo, Q. Jiang, J.-H. Chu, D. Xiao, X. Xu
Department of Physics, University of Washington, Seattle, Washington 98195, USA

S. Ha, S. Kim, S. Choi, H. Kim
Center for Ultrafast Phase Transformation, Department of Physics, Sogang University, Seoul 04107, Korea

X.-W. Zhang, T. Cao, D. Xiao, X. Xu
Department of Materials Science and Engineering, University of Washington, Seattle, Washington 98195, USA

S. Chun, J. Park, I. Eom
Pohang Accelerator Laboratory, POSTECH, Pohang, Gyeongbuk 37673, Korea

Z. Yang, E. J. G. Santos
Institute for Condensed Matter and Complex Systems, School of Physics and Astronomy, The University of Edinburgh, Edinburgh EH9 3FD, United Kingdom

E. J. G. Santos
Donostia International Physics Center, Donostia-San Sebastian, 20018, Spain

K. Oh, Y. Su, N. Gedik
Department of Physics, Cambridge, Massachusetts Institute of Technology, Massachusetts 02139, USA

A. Zong
Departments of Physics and Applied Physics, Stanford University, Palo Alto, California 94305, USA

# Table of contents

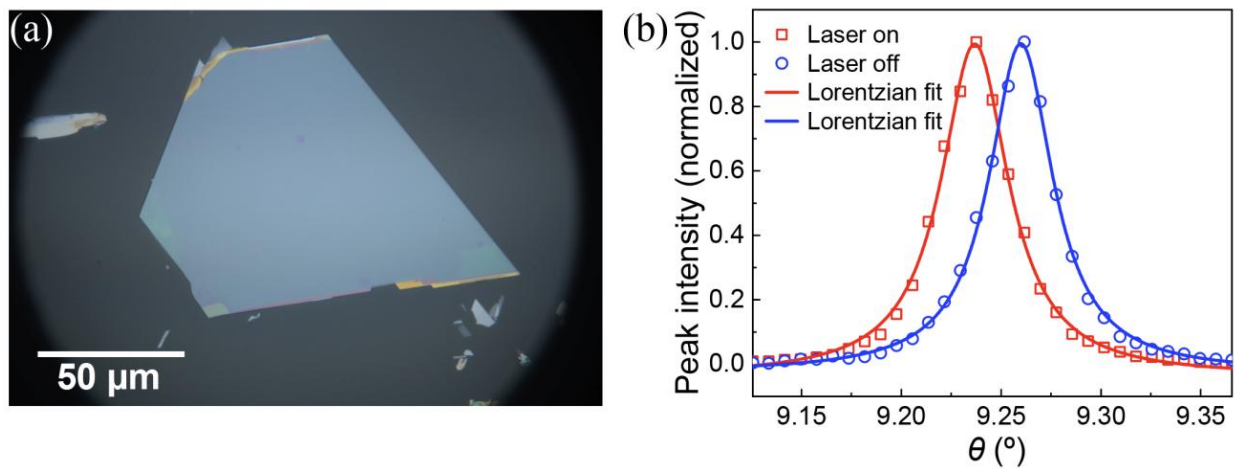

**Figure S1.** Optical image and rocking scan of FePS$_3$ sample. (a) Optical image of the 195.8-nm-thick sample flake for the experiment. (b) Rocking scan of 002 Bragg peak at +142 ps time delay (laser on) and before laser excitation (laser off) with 2.6 mJ cm$^{-2}$ incident fluence at 296 K. Solid lines are Lorentzian fits to the data.

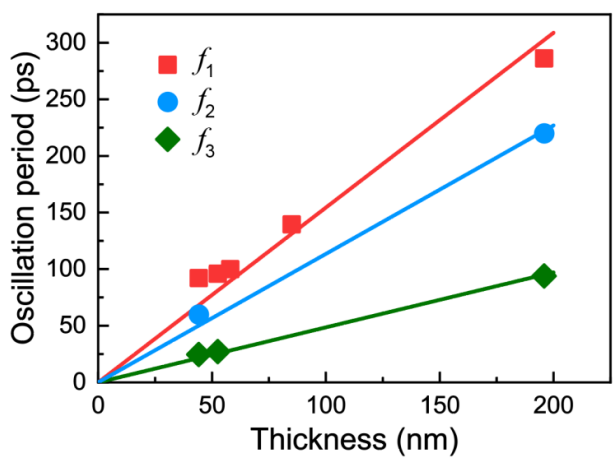

**Figure S2.** Thickness-dependent oscillation periods of the three modes. Each thickness data point corresponds to an independent sample flake with uniform thickness. Solid lines are linear fits crossing the origin.

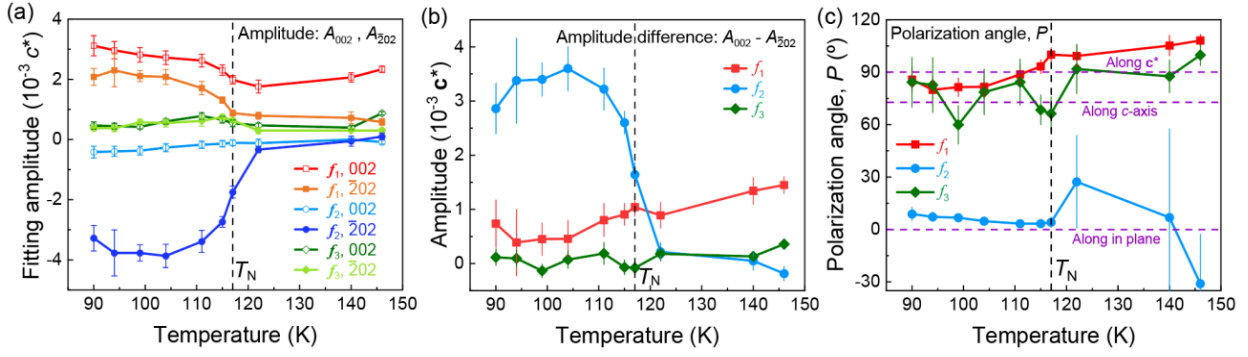

**Figure S3.** (a) Temperature-dependent oscillation amplitudes from the three-component fitting for both 002 and $\bar{2}02$ peaks. (b) The differential amplitude ($A_{002} - A_{\bar{2}02}$) that indicates the shear component for each mode based on the data in (a). (c) Temperature-dependent phonon polarization angle $P$ (i.e., the direction of atomic displacement vector **u**) based oscillation amplitude, see Section 3 in Supporting Information for details. The large error bars for the $f_2$ mode above $T_N$ are due to the vanishing $f_2$ amplitudes along both the in-plane and out-of-plane directions.

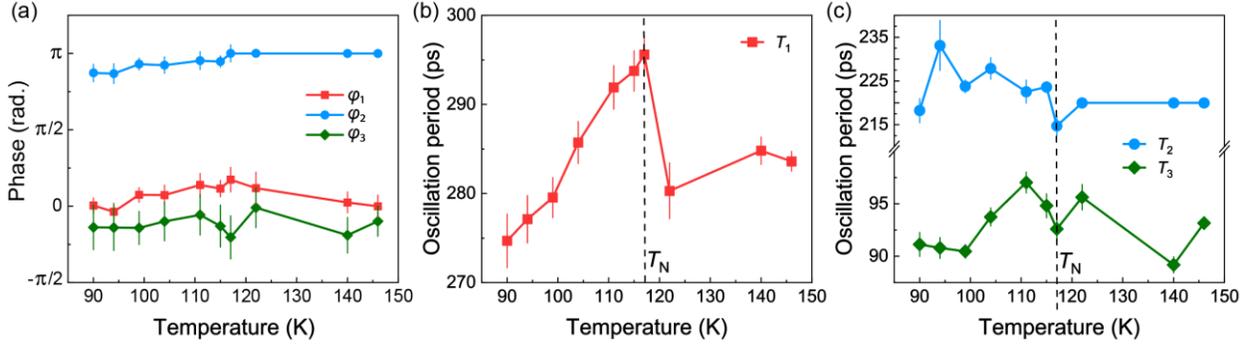

**Figure S4.** Temperature-dependent phases and sound velocities. (a) Temperature-dependent phases from the time-domain fitting. Each curve is the average of 002 and $\bar{2}02$ peak. (b)-(c) Temperature-dependent oscillation periods ($T_1$, $T_2$, $T_3$) from the time-domain fitting for $f_1$ mode (b), $f_2$ and $f_3$ modes (c). The data for each curve are the average of 002 and $\bar{2}02$ peak fitting. For the three data points above $T_N$ for $f_2$ mode, the amplitude is too weak for reliable fitting for $f_2$ component. Therefore, the oscillation period is fixed at 220 ps so that the amplitudes in Figure S3 can be extracted.

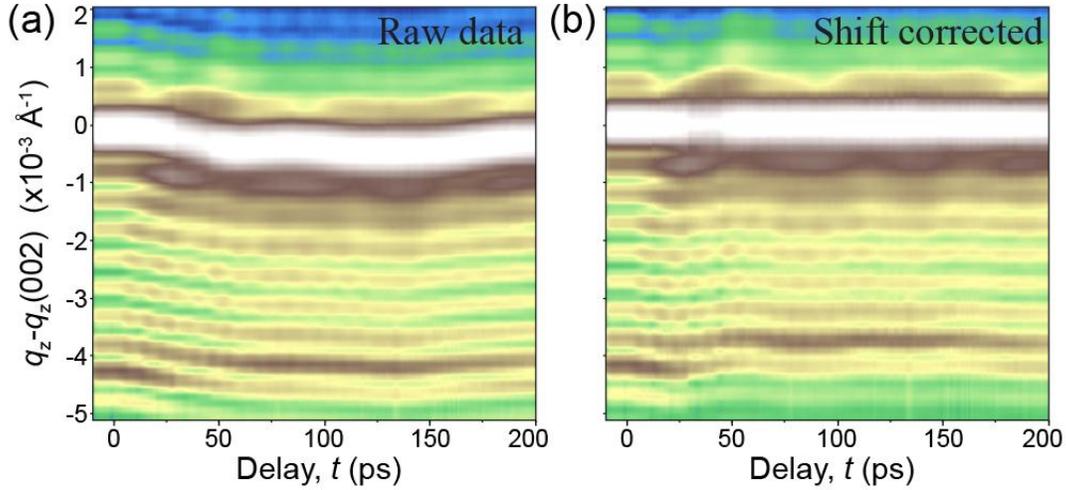

**Figure S5.** Shift correction of the 002 peak to eliminate the fundamental frequency of the acoustic oscillation due to the film thickness. (a) Momentum-resolved raw-data dynamics of 002 peak at room temperature. Data reproduced from Figure 3b. (b) Shift correction of data in (a) based peak center-of-mass shift of the main Bragg peak, which FFT is shown in Figure 3c. Other dispersion plots (Figure 3f and 3i) are obtained using the same shift-correction procedures as the 002 peak exemplified here.

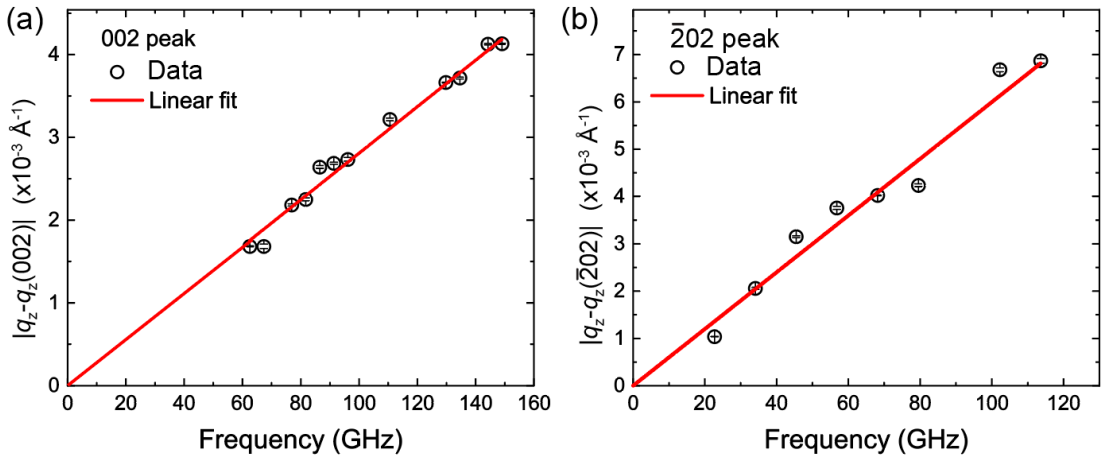

**Figure S6.** Linear fitting of the dispersion. (a) Linear fitting of the dispersion for 002 Bragg peak at room temperature, corresponding to the data shown in Figure 3c. The fitted slope of $2.808 \times 10^{-4}$ Å$^{-1}$ GHz$^{-1}$ corresponds to a sound velocity of 3561 m s$^{-1}$ for the $f_3$ mode. (b) Linear fitting of the dispersion for $\bar{2}02$ Bragg peak at 99 K, corresponding to the data shown in Figure 3f. The fitted slope of $5.995 \times 10^{-4}$ Å$^{-1}$ GHz$^{-1}$ corresponds to a sound velocity of 1668 m s$^{-1}$ for the $f_2$ mode.

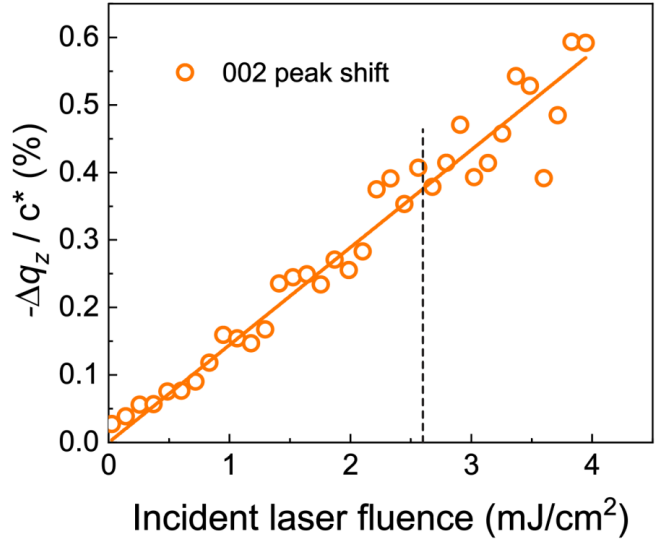

**Figure S7.** Laser fluence scan of the 002 peak at +142 ps time delay and 90 K. The vertical dashed line marks the 2.6 mJ cm$^{-2}$ incident laser fluence used in the work, unless specified otherwise.

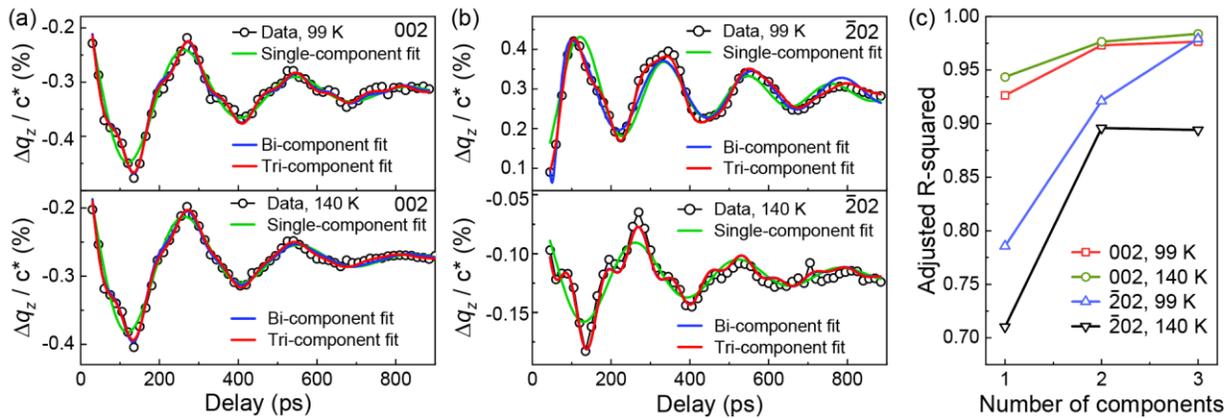

**Figure S8.** Comparison of one-, two-, and three-component fittings. (a) Fitting of the 002 peak dynamics at 99 K (top panel) and 140 K (bottom panel). Single-component, bi-component, and tri-component fitting functions are used for the fitting. (b) Fitting of the $\bar{2}02$ peak dynamics at 99 K (top panel) and 140 K (bottom panel). In the bottom panel, the blue curve (bi-component fit) is overlapped and behind the red curve. (c) Summary of the adjusted $R$-squared values for the fittings in (a) and (b).

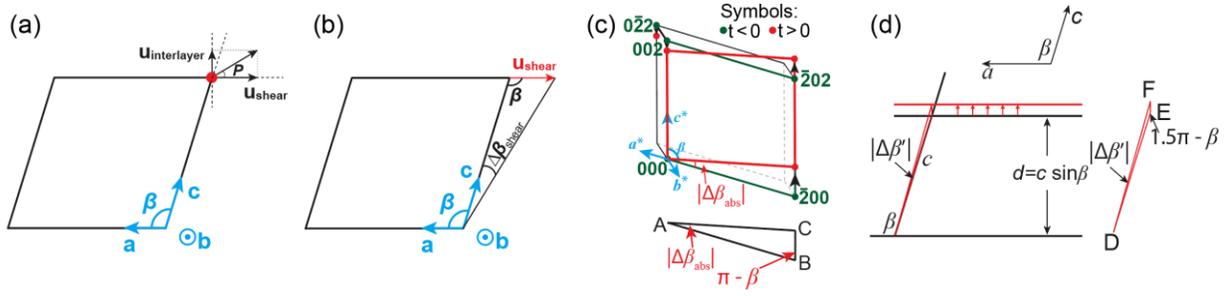

**Figure S9.** Illustration of phonon atomic displacement polarization. (a) Illustration of how the polarization angle $P$ is calculated based on the in-plane and out-of-plane atomic displacements. (b) Illustration of how the in-plane shear displacement $u_{shear}$ is extracted. (c) and (d) are reproduced from the Supplementary Figure 5 of Supplementary Reference [3]. (c) Illustration of how the absolute change of the monoclinic angle $\Delta\beta_{abs}$ is extracted. (d) Illustration of interlayer spacing-induced effective monoclinic angle change $\Delta\beta'$.

**Section 1.** Estimate of the sample temperature rise

The laser incident fluence is calculated based on the fluence at normal incidence times a factor of $\sin(\alpha)$, where $\alpha$ is the laser incident angle relative to the sample surface. Given the diffraction geometry and the 10° laser-x-ray crossing angle, the laser incident angles when measuring the 002 and $\bar{2}02$ peaks are 15.24° and 19.97°, respectively. The measurements at the two peaks were performed at the same laser power, which means that the incident laser fluence at the $\bar{2}02$ peak is a factor of 1.65 larger than that of the 002 peak. Since the 002 peak dynamics scales linearly with laser fluence (Figure S7), the $\Delta q_z$ peak shift in the main text for the 002 peak is rescaled by multiplying a factor of 1.65 so that the rescaled peak shifts measured at the 002 and $\bar{2}02$ peak are induced by the same pump fluence. The incident fluence $F_{in} = 2.6$ mJ cm$^{-2}$ for the 002 peak shown in Figure S7 has already included such a correction factor.

At the sample surface, the absorbed energy density ($E$) can be calculated by: $E = \frac{F_{in}(1-R)}{\delta}$, where $R \approx 0.25$ is the reflectivity for 400 nm photon at the designated incident angle, and $\delta \approx$ 95 nm is the optical penetration depth for 400 nm photon.[1] With these input parameters, we get $E = 205.3$ J cm$^{-3}$ = $1.21 \times 10^5$ J mol$^{-1}$. Given that the photon energy is 3.1 eV and the unit cell (not considering the unit cell doubling by the magnetic order) volume is 0.39 nm$^3$, the absorbed photon density at the top surface is $\rho_{h\nu} = \frac{E}{h\nu} = 0.16$ photons per unit cell. Assuming no heat dissipation to the sample environment in the first few picoseconds and quasi-thermalization between the lattice, electron, and spin at the surface, we can estimate the sample-surface temperature rise $\Delta T = T_f - T_i$ by integrating the specific heat: $E = \int_{T_i}^{T_f} C(T) dT$, where $T_i$ and $T_f$ are the initial and final sample temperatures, respectively, and $C(T)$ is the specific heat from the reference.[2] The electron-phonon coupling time constant is ~1 ps based on the Debye-Waller effect from the ultrafast electron diffraction measurements.[1] At an initial temperature of 90 K, we estimate a transient temperature rise at the sample surface to be 124 K.

**Section 2.** Time-domain fitting

To evaluate the proper fitting function (particularly the number of exponentially decaying sinusoidal components), we tried single-component, bi-component, and tri-component fittings. As shown in Figure S8a, for the 002 peak, both the bi- and tri-component functions fit the data better than single-component fitting regardless the sample is above and below $T_N$. The bi-component and tri-component fittings have similar adjusted $R$-squared values in the fitting (Figure S8c), suggesting that the third component (which is the $f_2$ mode), if it exists, has negligible amplitude in the 002 peak. For the $\bar{2}02$ peak fitting above $T_N$, as shown in Figure S8b bottom panel, the results are similar to the 002 peak, suggesting that the oscillation mostly contains $f_1$ and $f_3$ mode but negligible $f_2$ mode. By contrast, below $T_N$, the tri-component fitting is much better than the bi-component fitting, evidenced by a pronounced increase in adjusted R-squared value from the bi-component to tri-component fitting (Figure S8c). This is consistent with the fitted $f_2$ mode amplitude of $\bar{2}02$ peak in Figure 2c, where the $f_2$ mode amplitude dominates below $T_N$ but almost vanishes above $T_N$.

As the 002 peak exhibits weak temperature dependence in the dynamics, for three temperature points (94 K, 104 K, 115 K) that we didn't take delay scans for the 002 peak (see Figure 1c), we have used adjacent averaging with linear interpolation weighting to derive the dynamics at these three temperatures. For instance, the 94 K dynamic is weighted averaged based on the 90 K and 99 K data, with the weighting factor based on the distance of 94 K from the two temperatures. Then we perform the same time-domain fitting procedures as the other temperature points and compare the fitting results with the $\bar{2}02$ peak (Figure 2c and 2d).

In the time-domain fitting, to demonstrate that $f_2$ mode is negligible in the 002 peak, we fitted the 002 peak with a model accounting for only the $f_1$ and $f_3$ modes. The two-component fitting only led to a negligible difference in the adjusted $R^2$ value (<0.7%, Figure S8c) compared to the three-component fitting, indicating that $f_2$ mode oscillation is negligible in the 002 peak. By comparison, for the $\bar{2}02$ peak below $T_N$, the adjusted $R^2$ value of the fitting increases by 6% (from 92% to 98%) when comparing the two-component fit to the three-component fit, indicating the emergence of the third component (i.e., the $f_2$ mode) below $T_N$. Note that for the $\bar{2}02$ peak above $T_N$, a two-component fitting including $f_1$ and $f_3$ modes fits the data well and the contribution from $f_2$ mode is negligible, similar to the case of 002 peak. The amplitude of the $f_2$ mode in the $\bar{2}02$ peak below $T_N$ is the largest among all three modes. Moreover, the phases of the $f_1$ and $f_3$ modes are around zero, indicating a displacive excitation mechanism, while the phase of the $f_2$ mode is about $\pi$. The emergence and $\pi$-out-of-phase behavior of the $f_2$ mode leads to the overall oscillation sign flip across $T_N$, as shown in Figure 1e.

**Section 3.** Extraction of atomic displacement polarization angle

The angle of polarization $P$ here is defined relative to the in-plane $a$-axis, as shown in Figure 1b. The atomic displacement vector **u** for a specific phonon mode can be projected onto the in-plane ($u_{shear}$) and out-of-plane ($u_{interlayer}$) components. As shown in Figure S9a, the phonon polarization angle can be calculated as: $P = \tan^{-1}\frac{u_{interlayer}}{u_{shear}}$.

To calculate the out-of-plane atomic displacement $u_{interlayer}$, the longitudinal strain from the oscillation can be expressed as: $\frac{u_{interlayer}}{c\sin\beta} = -\frac{A_{002}}{2c^*}$, where $c\sin\beta$ is interlayer spacing, $A_{002}$ is the oscillation amplitude for the 002 peak, and $c^*$ is the reciprocal space vector. A a result, we obtain $u_{interlayer} = -\frac{A_{002}}{2c^*}c\sin\beta$.

To calculate the in-plane atomic displacement $u_{shear}$, the schematic drawing is shown in Figure 9b. Based on the Sine Theorem, $u_{shear} = -\frac{\sin(\Delta\beta_{shear})}{\sin(\pi-\beta-\Delta\beta_{shear})}c$, where $\beta = 107.2°$ is the monoclinic tilt angle, $c$ is the lattice constant, $\Delta\beta_{shear}$ is the interlayer shear angle that we will elaborate in the following. The sign of $u_{shear}$ is positive when pointing towards the +**a** direction (the corresponding $\Delta\beta_{shear}$ is negative). For a monoclinic structure as in the case of FePS$_3$, an interlayer expansion along the plane-normal **c*** direction would effectively reduce the monoclinic tilt angle, though there is no interlayer shear motion that we commonly refer to. Therefore, here we distinguish the monoclinic angle change by three concepts: (1) $\Delta\beta_{abs}$ is the absolute or total $\beta$ angle change; (2) $\Delta\beta'$ is the interlayer spacing change-induced effective $\beta$ angle change; (3) $\Delta\beta_{shear}$ is the angle change that is related to the interlayer shear displacement $\Delta\beta_{shear} = \Delta\beta_{abs} - \Delta\beta'$. As shown by Figure S9c, $\Delta\beta_{abs}$ can be resolved in the triangle $ABC$ (in which $AB = 2a^*$, $BC = \Delta q_{002} - \Delta q_{\bar{2}02} = A_{002} - A_{\bar{2}02}$ with $\Delta q_{002}$ and $\Delta q_{\bar{2}02}$ being the 002 and $\bar{2}02$ peak shift along the $c^*$, respectively; $A_{002}$ and $A_{\bar{2}02}$ being the fitted amplitudes of a specific phonon mode): $\Delta\beta_{abs} = \sin^{-1}(\frac{BC}{\sqrt{AB^2+BC^2-2*AB*BC*\cos(\pi-\beta)}}\sin\beta)$. With small-angle approximation $\Delta\beta_{abs} \approx \frac{BC}{\sqrt{AB^2+BC^2+2*AB*BC*\cos\beta}}\sin\beta = cc_1 * (A_{002} - A_{\bar{2}02})$, where $\Delta\beta_{abs}$ is in units of degrees, $cc_1 = \frac{\sin\beta}{\sqrt{AB^2+BC^2+2*AB*BC*\cos\beta}} \approx \frac{1}{0.043c^*}$ is a constant number. Therefore $\Delta\beta_{abs} \approx \frac{A_{002}-A_{\bar{2}02}}{0.043c^*}$, with $\Delta\beta_{abs}$ in units of degrees and the numerator being the amplitude difference plotted in Figure S3b. As Figure S9d shows, $\Delta\beta'$ can be resolved in the triangle $DEF$ (in which $DE = c$, $EF$ is calculated based on the strain relationship $\frac{EF}{d} = \frac{A_{002}}{q_{002}}$: $EF = c\sin\beta * \frac{A_{002}}{2c^*}$, $DF$ is calculated based on the Cosine Theorem $DF = \sqrt{EF^2 + DE^2 - 2 * EF * DE * \cos(1.5\pi - \beta)}$). Based on the Sine Theorem, $\Delta\beta' = \sin^{-1}(\frac{EF}{DF}\sin(3\pi/2 - \beta))$. With small-angle approximation $\Delta\beta' \approx \frac{EF}{DF}\sin\left(\frac{3\pi}{2} - \beta\right) = cc_2 * \frac{A_{002}}{c^*}$, where $\Delta\beta'$ is in units of degrees, $A_{002}$ is in relative units of $c^*$ plotted in Figure S3a, and $cc_2 \approx 8.0$ is a constant.

In summary, the phonon polarization angle $P = \tan^{-1}\frac{u_{\text{interlayer}}}{u_{\text{shear}}}$ is a function of the phonon oscillation amplitudes $A_{002}$ and $A_{\bar{2}02}$, both of which were from the time-domain fittings. Raw data of $A_{002}$ and $A_{\bar{2}02}$, as well as their differences, are plotted in Figure S3. The phonon polarization angles $P$ are plotted in Figure S3c and the main Figure 2.

**Section 4.** Sound velocity for the $f_2$ and $f_3$ modes

It is worth noting that the sound velocity of the interlayer shear ($f_2$) mode is 33% larger than 1120 m s$^{-1}$ previously measured by ultrafast electron diffraction and microscopy.[1] We first eliminated the trivial technical issues, such as inaccurate calibration of the delay time in these measurements. Then, we discuss two important differences between these measurements. The first is that the samples used in ultrafast electron diffraction measurements are thinner (~40 nm) than the sample (~200 nm) measured in this work, as thinner samples are required for electron transparency. As a result, possible surface oxidation[4] has a larger impact on the thinner samples. The second difference is that in thick samples, the optical excitation only directly excites part of the sample across $T_N$ due to the finite penetration depth of the excitation light. The thermal transport model (Section 5 in Supporting Information) captures the evolution of the lattice and spin temperatures (Figure 4c and 4d) across the sample, in which thermal gradients are presented. Since the sample is in a mixed state of antiferromagnet and paramagnet after excitation in thick samples, but mostly a paramagnetic state in thinner samples, the speed of sound for the interlayer shear mode is expected to be higher in the antiferromagnetic state. As the magnetization state of the sample still evolves in the thick sample during the measured time window (Figure 4d), the results measured in the thick sample are thus more complicated to model.

The discrepancy in sound velocity for the $f_3$ mode measured by the peak shift (Figure 1 and 2) and the dispersion relation (Figure 3) might be due to the uncertainty of the effective propagation length $l$, which can be different from the measured sample thickness $d$. The wavevector of the $f_3$ mode may not be exactly along the **c*** direction. Any deviation of the wavevector from the **c*** direction can increase the effective propagation distance $l$, thus improving the agreement of the two measurements.

**Section 5.** Computation of thermal transport dynamics

The simulation utilizes an extended version of a Microscopic Three-Temperature Model (M3TM) to compute the ultrafast laser-induced magnetization and temperature dynamics in FePS$_3$.

The model consists of two components: a modified Two-Temperature Model (2TM) describing heat transfer, and a microscopic spin-flip mechanism driven by electron-phonon scattering events that governs magnetization dynamics.

In the 2TM component, laser absorption with penetration depth $\lambda$ is modeled as

$$S(z,t) = \frac{S_0}{\lambda} e^{-\frac{(t-t_0)^2}{2\sigma^2}} e^{-z/\lambda}$$

using Lambert-Beer's law to account for the depth-dependent profile along the out-of-plane ($z$) direction. The temporal profile of the laser pulse is represented by a Gaussian function, characterized by a peak at time $t_0$ and a pulse duration $\sigma$. Laser-excited electrons quickly thermalize due to Coulomb scattering, resulting in a sharp rise in electronic temperature $T_e$. Energy is then transferred from the electron to the lattice through electron-phonon coupling, with coupling constant $g_{e-p}$, leading to equilibration with the phonon temperature $T_p$. The energy associated with magnetism is defined as $\dot{Q}_{e-s} = Jm\dot{m}/V_{at}$ with mean atomic volume $V_{at}$ and the mean field approximation exchange energy $J = 3\frac{S^2}{S(S+1)} k_B T_C$, and $m$ is the magnetization.

The full 2TM framework describes two coupled thermal baths:

$$C_e \frac{dT_e}{dt} = \frac{\partial}{\partial z}\left(\kappa_e \frac{T_e}{T_p}\frac{\partial T_e}{\partial z}\right) + g_{e-p}(T_p - T_e) + S(z,t) + \dot{Q}_{e-s}$$
$$C_p \frac{dT_p}{dt} = \frac{\partial}{\partial z}\left(\kappa_p \frac{\partial T_p}{\partial z}\right) - g_{e-p}(T_p - T_e)$$

The electron and phonon heat capacities $C_e$ and $C_p$ are described using the Sommerfeld approximation $C_e = \gamma_e T_e$ and the Einstein model $C_p = C_{p\infty}\frac{T_{Ein}^2}{T_p^2}\frac{\exp\frac{T_{Ein}}{T_p}}{\left(\exp\frac{T_{Ein}}{T_p}-1\right)^2}$, respectively. For simplicity, the material is modeled as a series of 2 nm-thick micro-cells (layers), where temperature gradients $\frac{\partial T_{e,p}}{\partial z}$ are approximated by the discrete temperature differences between adjacent layers. Thermal transport is governed by both electronic and phononic out-of-plane thermal conductivities ($\kappa_e$ and $\kappa_p$). At longer timescales, as the electron and phonon baths reach thermal equilibrium, heat diffusion is dominated by phonon transport across the heterostructure.

The magnetization dynamics are described by spin-flip processes resulting from electron-phonon scattering events[5,6] as follows:

$$\frac{dm}{dt} = -\frac{1}{S}\sum_{m_s=-S}^{m_s=+S} m_s \frac{df_{m_s}}{dt}$$

$$\frac{df_{m_s}}{dt} = -(W^+_{m_s} + W^-_{m_s})f_{m_s} + W^+_{m_s-1}f_{m_s-1} + W^-_{m_s+1}f_{m_s+1}$$

$$W^{\pm}_{m_s} = R\frac{Jm}{4Sk_B T_c}\frac{T_p}{T_c}\frac{e^{\mp\frac{Jm}{2Sk_B T_e}}}{\sinh\left(\frac{Jm}{2Sk_B T_e}\right)}(S(S+1) - m_s(m_s \pm 1))$$

The spin transitions are described using Boltzmann rate occupations $f_{m_s}$ of the $S_z$ component $m_s$. The dimensionless function $W^{\pm}_{m_s}$ represents the transition between neighboring spin states and is determined by the rate parameter $R$, defined as:[7]

$$R = 8\frac{a_{sf}g_{e-p}T_C^2 V_{at}}{\mu_{at}k_B T_{Debye}^2},$$

where the spin-flip probability $a_{sf}$ is an estimated microscopic parameter, and $V_{at}$ and $\mu_{at}$ are the atomic volume and magnetic moment per atom, respectively. The phonon temperature and depth-resolved demagnetization dynamics are summarized in Figure 4c and 4d.

Parameters:

| FePS$_3$ | Description | Value | Unit | Supplementary References |
|---|---|---|---|---|
| $\lambda$ | Penetration depth | $95 \times 10^{-9}$ | m | |
| $T_{Debye}$ | Debye temperature | 236 | K | [8] |
| $V_{at}$ | Magnetic atomic volume | $1.034 \times 10^{-28}$ | m$^3$ | |
| $\kappa_p$ | Phononic heat diffusion constant | 0.85 | W/m/K | [9] |
| $\kappa_e$ | Electronic heat diffusion constant | 0.0013 | W/m/K | |
| $\gamma_e$ | Sommerfeld constant of electronic heat capacity | 1500 | J/m$^3$/K$^2$ | |
| $C_{p\infty}$ | Maximal phononic heat capacity | $1.25 \times 10^6$ | J/m$^3$/K | [10] |
| $g_{e-p}$ | Electron-phonon coupling constant | $1.2 \times 10^{17}$ | W/m³/K | |
| Spin | Effective spin of the material | 2 | | |
| $T_C$ | Neel temperature | 120 | K | [12] |
| $\mu_{at}$ | Atomic magnetic moment | 5 | $\mu_B$ | [12,13] |
| $a_{sf}$ | Electron-phonon-scattering induced spin flip probability | 0.04 | | |

This model was originally developed for metallic materials but was soon adapted for nonmetallic two-dimensional magnetic semiconductors, such as $CrI_3$[14] and $Cr_2Ge_2Te_6$.[15] As the photoinduced carriers create a transient quasi-metallic state, making it possible to describe electron and lattice bath with effective temperatures $T_e$ and $T_p$.

**Supplementary References**